\documentclass[twocolumn,showpacs,preprintnumbers,amsmath,amssymb]{revtex4}
\usepackage[dvips]{graphicx}
\usepackage{color}

\def\ket#1{  \left\vert  #1   \right\rangle   }
\def\bra#1{  \left\langle  #1   \right\vert   }

\begin{document}

\title{Defeating entanglement sudden death by a single local
filtering}

\author{Michael Siomau and Ali A. Kamli}

\affiliation{Physics Department, Jazan University, P.O.~Box 114,
45142 Jazan, Kingdom of Saudi Arabia}

\date{\today}

\begin{abstract}
Genuine multipartite entanglement of a quantum system can be
partially destroyed by local decoherence. Is it possible to retrieve
the entanglement to some extent by a single local operation? The
answer to this question depends very much on the type of initial
genuine entanglement. For initially pure W and cluster states and if
the decoherence is given by generalized amplitude damping, the
answer is shown to be positive. In this case, the entanglement
retrieving is achieved just by redistributing the remained
entanglement of the system.
\end{abstract}


\maketitle

Being a cornerstone for quantum technologies, entanglement remains
maybe the most mysterious feature of quantum theory. Despite many
remarkable results devoted to quantifying \cite{Horodecki:09} and
measuring \cite{Guehne:09} entanglement as well as to describing its
dynamical change \cite{Lopez:08,Konrad:08,Farias:09,Kim:12}, the
deep understanding of fundamental laws of entanglement evolution,
and especially for complex multipartite systems, is still
challenging. Nevertheless, it is known for certain that entanglement
of a quantum system is very fragile and can be seriously harmed by
interaction of the system with its environment
\cite{Aolita:08,Borras:09,Aolita:09,Siomau:10}. If, moreover,
different parts of the system interact with local environments, the
entanglement can be completely destroyed in a finite time, although
coherence of the system is lost asymptotically \cite{Yu:06}. This
effect, which is known today as entanglement sudden death (ESD),
establishes serious limitations on efficient practical utilization
of entanglement in communication and computing \cite{Almeida:07}.
Therefore, it is highly desired to find a way to reverse ESD by
local manipulations when it is possible.

For bipartite entangled systems, reversing ESD by any local
manipulations is impossible \cite{Sun:10}, unless one is able to
control local environment of the system \cite{Lopez:08} or acquire
information about the whole system prior to its interaction with the
environment \cite{Kim:12}. In this paper we show that, for systems
exhibiting certain types of multipartite entanglement, reversing ESD
may be successful probabilistically and without mentioned additional
assumptions. Focusing on qubits, the two-dimensional quantum
systems, we shall analyze entanglement dynamics of different locally
unitary inequivalent entangled states, such as W,
Greenberger-Horne-Zeilinger (GHZ) \cite{Duer:00} and cluster
\cite{Briegel:01} states. We shall only assume that a multiqubit
system is partially affected by decoherence, i.e. some qubits are
preserved from detrimental influence of environment. The consequence
of this assumption is that entanglement among decohering qubits is
lost much faster comparing to the entanglement between decohering
and non-decohering qubits. This makes possible to retrieve genuine
multiqubit entanglement after partial ESD by redistributing the
entanglement which remained in the system. To achieve such
redistribution we request that just a single non-decohering qubit is
available for a local operation. This is indeed the minimal demand
on accessibility of the multiqubit system in practice.

As noise model for decoherence we chose generalized amplitude
damping (GAD) -- a reservoir in thermal equilibrium with a qubit at
a finite temperature \cite{Nielsen:00}. The GAD can be represented
in terms of four (Kraus) operators
\begin{eqnarray}
 \label{gad}
 K_1 &=& \sqrt{1-p} \left( \begin{array}{cc} 1 & 0 \\ 0 &
\sqrt{1-\gamma} \end{array} \right) , \,
 K_2 = \sqrt{1-p} \left( \begin{array}{cc} 0 & \sqrt{\gamma} \\
0 & 0 \end{array} \right) \, ,
 \nonumber  \\[0.1cm]
 K_3 &=& \sqrt{p} \left( \begin{array}{cc} \sqrt{1-\gamma} & 0 \\ 0
& 1 \end{array} \right) , \,
 K_4 = \sqrt{p} \left( \begin{array}{cc} 0 & 0 \\ \sqrt{\gamma} & 0
\end{array} \right) \, .
\end{eqnarray}
where the dynamical parameter $\gamma$ can be expressed as $1 -
e^{\Gamma t}$ through the coupling constant $\Gamma$ (which defines
the temperature of the reservoir, for instance) and the time of
interaction $t$. In order to reduce number of involved parameters we
shall later assume $p \equiv 1/2$, although our results remain true
without this latter assumption.

\begin{figure}[b]
\begin{center}
\includegraphics[scale=0.6]{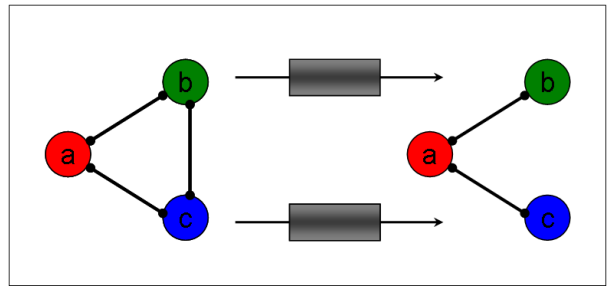}
\caption{(Color online) If two qubits {\rm b} and {\rm c} of a
three-qubit system, which is initially prepared in W state
(\ref{W3}), undergo GAD simultaneously, ESD occurs between these
qubits faster than between qubits {\rm a} and {\rm b} ({\rm a} and
{\rm c}). A local filtering on the qubit {\rm a} can retrieve
entanglement between qubits {\rm b} and {\rm c} with some
probability.}
 \label{fig-1}
\end{center}
\end{figure}

Let us start our discussion with the maximally entangled three-qubit
W state which can be written in a computational basis as
\begin{equation}
 \label{W3}
\ket{W_3} = \frac{1}{\sqrt{3}} \left( \ket{0_a0_b1_c} +
\ket{0_a1_b0_c} + \ket{1_a0_b0_c} \right) \, .
\end{equation}
Suppose, two qubits (let say {\rm b} and {\rm c}) of the three qubit
state undergo GAD simultaneously as drawn in Fig.~\ref{fig-1}.
Following quantum operation formalism \cite{Nielsen:00}, the final
state of the system is given by
\begin{equation}
 \label{sum-represent}
 \rho_{\rm fin} = \sum_{ij} S_{ij} \, \rho_{\rm ini} \, S_{ij}^\dag \, ,
\end{equation}
where $S_{ij} = I \otimes K_i \otimes K_j$ for $i,j=1..4$, $I$ is
the identity matrix, $\rho_{\rm ini} = \ket{W_3}\bra{W_3}$ and the
condition $\sum_{ij} S_{ij}^\dag \, S_{ij} = I$ is fulfilled.

\begin{figure}
\begin{center}
\includegraphics[scale=0.9]{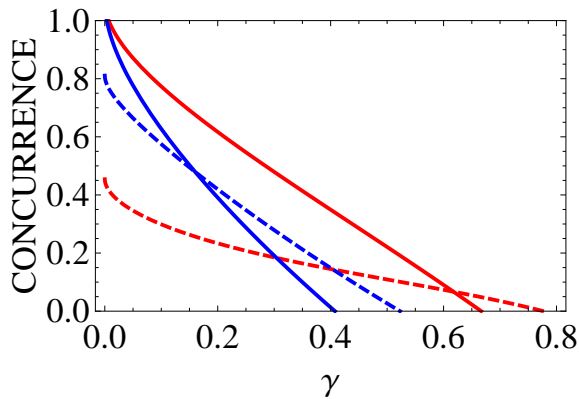}
\caption{(Color online) Concurrences $C_{\rm ab} = C_{\rm ac}$ (red)
and $C_{\rm bc}$ (blue) before (solid) and after (dashed) local
filtering. Entanglement between qubits {\rm b} and {\rm c} is
retrieved after ESD by the filtering.}
 \label{fig-2}
\end{center}
\end{figure}

Entanglement between any two qubits of the the three-qubit system
can be quantified with Wootters concurrence \cite{Wootters:98}
ignoring the third qubit. The concurrence is given by $C = {\rm max}
\{ 0, \; \lambda^1 - \lambda^2 - \lambda^3 - \lambda^4\}$, where
$\lambda^i$ are the square roots of the four nonvanishing
eigenvalues of the non-Hermitean matrix $\rho\: (\sigma_y \otimes
\sigma_y ) \rho^\ast (\sigma_y \otimes \sigma_y )$, if taken in
decreasing order. Because of the symmetry of the process in
Fig.~\ref{fig-1}, concurrences $C_{\rm ab}$ and $C_{\rm ac}$, which
are defined between corresponding qubits, are equal for all
parameters $\gamma$ of the GAD. At the same time, concurrence
$C_{bc}$ differs significantly from concurrences $C_{\rm ab}$ and
$C_{\rm ac}$ as shown in Fig.~\ref{fig-2}. When $C_{\rm bc}$ vanish
for $\gamma \approx 0.41$, each pair of qubits {\rm ab} and {\rm ac}
preserve significant amount of quantum information of $0.34$ ebit.

Let us now perform a filtering operation on qubit {\rm a}. This
operation can be written in a computational basis as
\begin{equation}
 \label{filtering}
F = \sqrt{1-\kappa} \ket{0}\bra{0} + \sqrt{\kappa} \ket{1} \bra{1},
\qquad 0<\kappa<1 \, .
\end{equation}
Filtering is a non-trace-preserving map which is known to be capable
of increasing entanglement with some probability \cite{Gisin:96}.
Practically, this map can be realized as a null-result weak
measurement \cite{Sun:10}.

\begin{figure}[b]
\begin{center}
\includegraphics[scale=0.6]{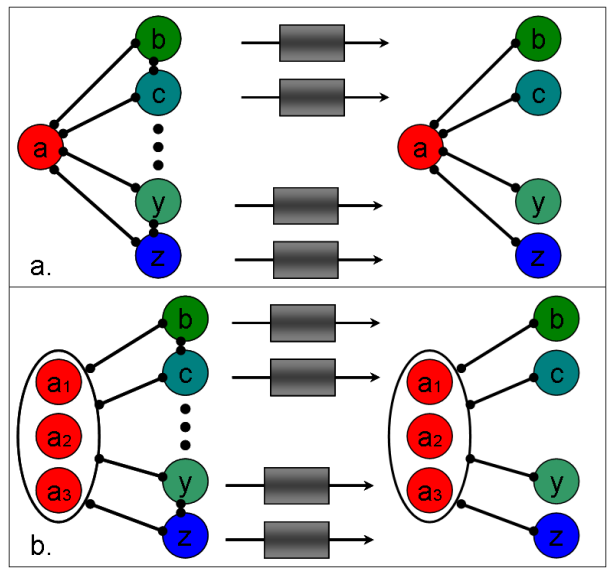}
\caption{(Color online) {\rm a.} Entanglement between qubits
$b,c...y,z$ can be retrieved after ESD by a single filtering, but
with a small probability. {\rm b.} This probability can be
increased, if $k$ qubits of the $N$-qubit W state are preserved from
decoherence and are the subject of filtering.}
 \label{fig-3}
\end{center}
\end{figure}

When filtering (\ref{filtering}) is applied to the qubit {\rm a},
the final three-qubit state $\left[{\rm F \otimes GAD \otimes
GAD}\right] \rho_{\rm ini}$ as well as the bipartite concurrences
$C_{\rm ab}$, $C_{\rm ac}$ and $C_{\rm bc}$ are dependent of two
parameters $\gamma$ and $\kappa$. For fixed $\gamma$, concurrences
$C_{\rm ab}$ and $C_{\rm ac}$ decrease with increasing $\kappa$,
while concurrence $C_{\rm bc}$ simultaneously increase. Let us
assume that $\gamma = 0.41$, then concurrence $C_{\rm bc}$ vanish
before filtering as we mentioned above. Demanding that after the
filtering all bipartite concurrences are equal, we can find
parameter $\kappa \approx 0.24$. For parameters $\gamma \approx
0.41$ and $\kappa \approx 0.24$, the entanglement between qubits
{\rm b} and {\rm c} is retrieved after ESD and $C_{\rm ab}=C_{\rm
ac}=C_{\rm bc} \approx 0.14$ ebit. This retrieving is, however,
probabilistic with probability $p=0.37$. The probability is
calculated from the norm of the final three-qubit state $\left[{\rm
F \otimes GAD \otimes GAD}\right] \rho_{\rm ini}$.

Our consideration can be generalized to the case of a $N$-qubit W
state. If $N-1$ qubits undergo GAD simultaneously as drawn in
Fig.~\ref{fig-3}{\rm a}, entanglement between the decohering qubits
can be retrieved after ESD by a single filtering. However, the
amount of retrieved entanglement as well as the probability of the
retrieving decrease rapidly with number of decohering qubits. This
situation can be changed dramatically, if $k>1$ qubits of the
$N$-qubit W state are preserved from decoherence as shown in
Fig.~\ref{fig-3}{\rm b}. In this case, one can increase
significantly both the amount of retrieved entanglement and the
probability of retrieving by performing filtering on the $k$ qubits
simultaneously. We observed significant increase of pairwise
entanglement among non-decohering qubits after the filtering, in the
case if $k>1$ qubits are preserved from decoherence and become the
subject of the filtering.

Having completed our analysis of W states and before moving to
cluster states we would like to note that genuine GHZ entanglement
can not be retrieved if ESD happened between two qubits of the
multiqubit GHZ state. In fact, if ESD appears between any two qubits
of the multiqubit GHZ state, all the qubits suddenly become
disentangled and no local operations can retrieve GHZ entanglement.

We are now at the position to consider entanglement dynamics of
cluster states under GAD. Because three-qubit cluster state is
locally unitary equivalent to the GHZ state, the simplest nontrivial
member of the family of cluster states is the four-qubit state
\cite{Briegel:01}
\begin{eqnarray}
 \label{cluster}
\ket{C_4} = \frac{1}{2} ( \ket{0_a0_b0_c0_d} &+& \ket{0_a0_b1_c1_d}
\nonumber \\[0.1cm]
 &+& \ket{1_a1_b0_c0_d} - \ket{1_a1_b1_c1_d} ) \, .
\end{eqnarray}
This state is less symmetric than four qubit GHZ and W states. For
the cluster state, only permutations of qubits {\rm ab} and {\rm cd}
leave the state unchanged, while for the GHZ and W states any two
qubits can be permuted without changing the state. The consequence
of the low symmetry of the cluster state is that it does not have
pairwise entanglement as W states: any two qubits are disentangled
if the other two qubits are discarded. At the same time, any three
qubits are entangled, if the remaining one is ignored. This differs
cluster states from GHZ states.

\begin{figure}
\begin{center}
\includegraphics[scale=0.6]{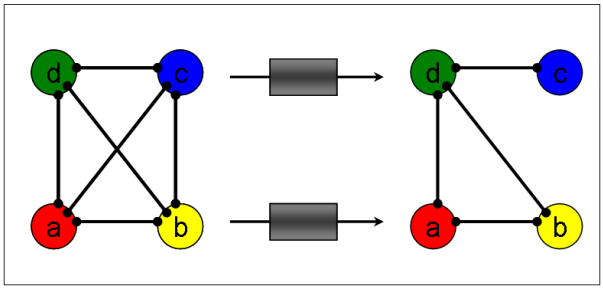}
\caption{(Color online) If two qubits {\rm b} and {\rm c} of the
cluster state (\ref{cluster}) are affected by GAD, ESD occurs first
between qubit {\rm c} and the pair {\rm ab}. Entanglement between
these qubits can be retrieved by a single local filtering on qubit
{\rm a}.}
 \label{fig-4}
\end{center}
\end{figure}

Since even pure cluster state (\ref{cluster}) does not contain
pairwise entanglement, Wootters concurrence is not suitable to
describe its entanglement dynamics. Another entanglement measure has
to be chosen to quantify entanglement of the cluster state properly.
We shall use a generalization of Wootters concurrence to bipartite
finite-dimensional systems \cite{Rungta:01,Ou:08}. The bipartite
concurrence of $d_1\otimes d_2$-dimensional system is given by $BC
\equiv \sqrt{\sum_{mn} C_{mn}^2}$ where $C_{mn} = {\rm max} \{ 0,
\lambda_{mn}^1 - \lambda_{mn}^2 - \lambda_{mn}^3 - \lambda_{mn}^4
\}$ and the $\lambda_{mn}^k, \, k=1..4$ are the square roots of the
four nonvanishing eigenvalues of the matrix $\rho\,
\tilde{\rho}_{mn}$, if taken in decreasing order. These matrices
$\rho\: \tilde{\rho}_{mn}$ are formed by means of the density matrix
$\rho$ and its complex conjugate $\rho^*$, and are further
transformed by the operators $\{ S_{mn} = L_m \otimes L_n,\; m =
1,...,M,\; n = 1,...,N \}$ as: $\tilde{\rho}_{mn} = S_{mn} \rho^\ast
S_{mn}$.  In this notation, moreover, $L_m$ and $L_n$ are generators
of groups SO$(M)$ and SO$(N)$ respectively, where $N=d_1(d_1-1)/2$
and $M=d_2(d_2-2)/2$.

In the following analysis we shall focus on bipartite concurrences
of the form $BC_{ij|k}$, where $i,j,k = a,b,c,d$ and $i\neq j\neq k
\neq i$. These bipartite concurrences quantify entanglement between
a single qubit $k$ and a pair of qubits $ij$ discarding the fourth
qubit of the cluster state (\ref{cluster}). If concurrence
$BC_{ij|k}$ vanishes, we say that ESD has occurred for the partition
$k|ij$, i.e. between the qubit $k$ and the pair of qubits $ij$.

\begin{figure}
\begin{center}
\includegraphics[scale=0.9]{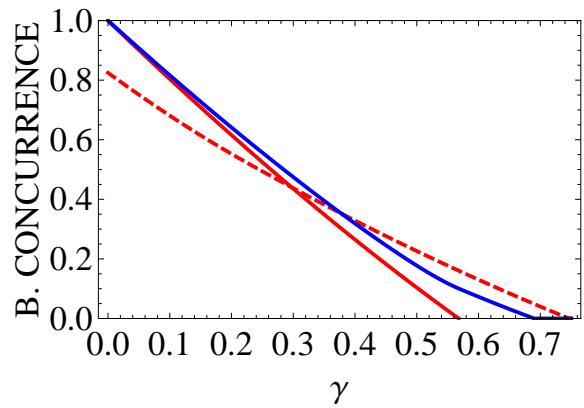}
\caption{(Color online) Bipartite concurrences $BC_{ab|c}$ (red) and
$BC_{cd|b}$ (blue) before (solid) and after (dashed) local filtering
on {\rm a}. Parameter of the filtering is chosen to be $\kappa=0.2$.
Remarkably, the bipartite concurrence $BC_{cd|b}$ does not change
after the filtering.}
 \label{fig-5}
\end{center}
\end{figure}

Let us assume as before that just two qubits of the four-qubit
cluster state (\ref{cluster}) undergo GAD simultaneously. It is
important to note that if qubits {\rm ab} or {\rm cd} are the
subject of GAD, the entanglement between different parts of the four
qubit system decreases asymptotically with parameter $\gamma$ and
ESD never occurs between partitions of the cluster state. This is
the consequence of the permutational symmetry of the cluster state
(\ref{cluster}). Nevertheless, if qubits {\rm bc} or {\rm ad} are
influenced by GAD, ESD is present in the system for some coupling
parameters $\gamma$. Let qubits {\rm b} and {\rm c} undergo GAD
simultaneously as shown in Fig.~\ref{fig-4}. The dynamics of the
cluster state (\ref{cluster}) is given by Eq.~(\ref{sum-represent})
with $S_{ij} = I \otimes K_i \otimes K_j \otimes I$ for $i,j=1..4$
and $\rho_{\rm ini} = \ket{C_4}\bra{C_4}$. By computing bipartite
concurrences for different partitions of the four-qubit system, we
have found that ESD appears first for the partition ${\rm ab|c}$ for
$\gamma \approx 0.57$ and later for partition ${\rm cd|b}$ for
$\gamma \approx 0.69$ as displayed in Fig.~\ref{fig-5}.

For $ 0.69 > \gamma \geq 0.57$, entanglement between qubits {\rm ab}
and {\rm c} can be retrieved by the local filtering on qubit {\rm a}
as shown in Fig.~\ref{fig-5}. Parameter of the filtering $\kappa$
can be chosen freely, unless we make additional demands for
entanglement retrieving as discussed below. The probability of the
entanglement retrieving is given as before by the norm of the final
state $\left[{\rm F \otimes GAD \otimes GAD \otimes I}\right]
\rho_{\rm ini}$.

It is important to note that the entanglement between qubits {\rm
cd} and {\rm b} is not influenced by the filtering, since the qubit
{\rm a} is discarded in the bipartite concurrence $BC_{cd|b}$. From
one hand, it means that if ESD occurred for partition ${\rm cd|b}$
(for $\gamma \geq 0.69$), entanglement between these qubits can not
be retrieved by the local filtering. Moreover, after ${\rm cd|b}$
entanglement is destroyed, the filtering can not retrieve ${\rm
ab|c}$ entanglement any more. From the other hand, invariance of
${\rm cd|b}$ entanglement with regard to the filtering on qubit {\rm
a} means that some bipartite entanglement in the system decreased
after the filtering. Indeed, there must be some price for retrieving
entanglement between qubits {\rm ab} and {\rm c} while ${\rm cd|b}$
entanglement remains the same. The price for the entanglement
retrieving is the decrease of ${\rm ad|b}$ and ${\rm ad|c}$
entanglement. In principle, one can set an additional condition for
the balance of entanglement in term of the bipartite concurrences
$BC_{ab|c}$, $BC_{ad|b}$ and $BC_{ad|c}$ in order to define the
parameter $\kappa$ of the filtering properly.

As we have seen, the analysis of the entanglement retrieving for the
four-qubit cluster state (\ref{cluster}) is much more complicated
than for $N$-qubit W states. This analysis involves computation of
twelve bipartite concurrences $BC_{ij|k}$ (for $i,j,k=a,b,c,d$ and
$i\neq j\neq k \neq i$) which, moreover, behave differently with
respect to each other because of the low symmetry of cluster state.
This makes difficult a straightforward generalization of our results
to the case on $N$-qubit cluster states for $N>4$ using approach
presented in this paper, i.e. by analyzing bipartite concurrences
between differen partitions of the state. Nevertheless, it seems
possible to use a single filtering for entanglement retrieving in
high order cluster states.

Our proposal for entanglement retrieving after ESD by a single local
filtering may find its applications in quantum communication and
quantum computing. It is known, in particular, that W states can be
used for communication between several remote recipients
\cite{Agraval:06}. Cluster states, in turn, are the resource for
measurement-based quantum computation \cite{Briegel:09}. In both
these applications, protection of genuine multipartite entanglement
against partial disentanglement is highly desired to ensure correct
and rigor quantum information processing.

In conclusion, we have shown that genuine multipartite entanglement
of W and cluster states, which has been partially lost due to
detrimental influence of local environments, can be
probabilistically retrieved to some extent by a single local
filtering. This retrieving is achieved only by redistributing
entanglement which remained in the system after the decoherence and
does not rely on ability to control environmental degrees of freedom
\cite{Lopez:08} or acquiring information about the whole system
before the decoherence \cite{Kim:12}. Therefore, we believe that our
scheme for entanglement retrieving can be successfully combined with
these alternative schemes.

\begin{acknowledgments}
We thank Mohammad Al-Amri and Suhail Zubairy for fruitful
discussions during our meetings in Jazan University and KACST. We
are also grateful to Luiz Davidovich for his comments.
\end{acknowledgments}

\end{document}